\documentclass[amsmath,amssymb,12pt]{article}
\usepackage{graphicx}
\usepackage{bm}
\begin{document}
\title{How many clusters? An information theoretic perspective}
\author{Susanne Still and William Bialek \\
Department of Physics, and\\
Lewis-Sigler Institute for Integrative Genomics\\
Princeton University \\
Princeton, New Jersey 08544, USA \\
{susanna,wbialek}@princeton.edu }
\date{}
\maketitle

\begin{abstract}
Clustering provides a common means of identifying structure in complex
data, and there is renewed interest in clustering as a tool for the analysis
of large data sets in many fields.  A natural question is how many clusters
are appropriate for the description of a given system.  Traditional
approaches to this problem are based either on a framework in which clusters of a
particular shape are assumed as a model of the system or on a two-step procedure in which 
a clustering criterion 
determines the optimal assignments for a given number of clusters and a separate 
criterion measures the goodness of the classification to determine the number of clusters.
In a statistical mechanics approach, clustering can be seen as a trade--off between 
energy-- and entropy--like terms, with lower temperature driving the proliferation
of clusters to provide a more detailed description of the data. 
For finite data sets, we expect that there is a limit to the meaningful structure 
that can be resolved and therefore a minimum temperature beyond which we will 
capture sampling noise. This suggests that correcting the clustering criterion
for the bias which arises due to sampling errors will allow us to find a clustering solution 
at a temperature which is optimal in the sense that we capture maximal meaningful 
structure --- without having to define an external criterion for the goodness or 
stability of the clustering. We show that, in a general information theoretic 
framework, the finite size of a data set determines an optimal temperature, and we
introduce a method for finding the maximal number of clusters which can be 
resolved from the data in the hard clustering limit.
\end{abstract}
\maketitle

\section{Introduction}
Much of our intuition about the world around us involves the idea of
clustering:  many different acoustic waveforms correspond to the same
syllable, many different images correspond to the same object, and so on.
It is plausible that a mathematically precise notion of clustering in
the space of sense data may approximate the problems solved by our brains.
Clustering methods also are used in many different scientific domains
as a practical tool to evaluate structure in complex data. Interest in
clustering has increased recently because of new areas of application,
such as data mining, image-- and speech--processing and bioinformatics.
In particular, many groups have used clustering methods to analyze the
results of genome--wide expression experiments, hoping to discover genes
with related functions as members of the same cluster; see, for example,
Eisen, Spellman, Brown and Botstein (1998).
A central issue in these and other applications of clustering is how many
clusters provide an appropriate description of the data. The estimation of the 
true number of classes has been recognized as ``one of the most difficult 
problems in cluster analysis'' by Bock (1996), who gives a review of some 
methods that address the issue.\\
\\
The goal of clustering is to group data in a meaningful way. This is achieved
by optimization of a so called ``clustering criterion'' (an objective function), and
a large variety of intuitively reasonable criteria have been used in the literature
(a summary is given in Gordon, 1999).
Clustering methods include agglomerative clustering procedures such as described by
Ward (1963) and iterative re-allocation methods, such as the commonly used K--means algorithm 
(Lloyd, 1957; MacQueen, 1967), which reduces the sum of squares
criterion, the average of the within cluster squared distances.
More recently, algorithms with physically inspired criteria were
introduced (Blatt, Wiseman and Domany 1996; Horn and Gottlieb, 2002). All these
clustering methods have in common that the number
of clusters has to be found by another criterion. Often, a two-step procedure is performed: 
The optimal partition is found for a
given data set, according to the defined objective function, and then a separate criterion
is applied to test the robustness of the results against noise due to finite sample size.
Such procedures include the definition of an intuitively reasonable criterion for the
goodness of the classification, as in Tibshirani, Walther and Hastie (2001), or performing 
cross--validation (Stone, 1974) and related methods in order to estimate the
prediction error and to find the number of clusters that minimizes this
error (e.g. Smyth, 2000). Roth, Lange, Braun and Buhmann (2002) quantify the goodness of the 
clustering via a resampling approach.  \\
\\ It would be attractive if these two steps could be combined in a single principle.
In a sense this is achieved in the probabilistic mixture model approach, but at the
cost of assuming that the data can be described by a mixture of $N_c$ multivariate
distributions with some parameters that determine their shape. Now the
problem of finding the number of clusters is a statistical model selection
problem. There is a trading between complexity of the model and
goodness of fit. One approach to model selection is to compute the total
probability that models with $N_c$ clusters can give rise to the data,
and then one finds that phase space factors associated with the
integration over model parameters serve to discriminate against more
complex models (Balasubramanian, 1997). This Bayesian approach has been
used to determine the number of clusters (Fraley and Raftery, 2002). \\
\\ From an information theoretic point of view, clustering is most
fundamentally a strategy for lossy data compression: the data are partitioned into
groups such that the data could be described in the most efficient way
(in terms of bit cost) by appointing a representative to each group. The clustering
of the data is achieved by compressing the original data into
representatives, throwing away information that is not relevant to the analysis.
While most classical approaches in statistics give an explicit definition
of a similarity measure, in rate--distortion theory we arrive at a notion of similarity
through a fidelity criterion implemented by a distortion function (Shannon 1948).
The choice of the distortion function provides an implicit distinction
between relevant and irrelevant information in the raw data. The notion
of relevance was made explicit by Tishby, Pereira and Bialek (1999) who defined
relevant information as the information which the data provide about an auxiliary variable
and performed lossy compression, keeping as much relevant information as possible.
This formulation, termed ``Information Bottleneck method'' (IB), is attractive, because the 
objective function follows only from information theoretical principles. In particular, 
this formulation does not require an explicit definition of a measure for similarity or distortion. 
The trade--off between the complexity of the model on one hand and the amount of relevant information it 
captures on the other hand is regulated by a trade-off parameter. For a given problem, the complete 
range of this trade--off is meaningful, and the structure of the trade--of characterizes the ``clusterability'' of the data.
However, for a {\it finite} data set, there should be a maximal value for this
trade--off after which we start to ``overfit,'' and this issue has not yet been addressed in the 
context of the IB. In related work, Buhmann and Held (2000) derived for a particular class of 
histogram clustering models a lower bound on the annealing temperature from a bound on the 
probability of a large deviation between the error made on the training data and the expected error.\\
\\ In this article, we follow the intuition that if a model --- which, in this context, is a
(probabilistic) partition of the data set --- captures information (or structure)
in the data, then we should be able to quantify this structure in a way that corrects automatically
for the overfitting of finite data sets. Attempts to capture only this ``corrected information''
will, by definition, not be sensitive to noise. Put another way, if we would separate at the outset
real structure from spurious coincidences due to undersampling, then we could fit only the real
structure. In the context of information estimation from finite samples there is a significant
literature on the problem and we argue here that the known finite sample correction to information
estimates is (in some limits) sufficient to achieve the ``one-step'' compression and clustering in
the sense described above, leading us naturally to a principled method of finding the best
clustering that is consistent with a finite data set.\\
\\ We should point out that in general we are not looking for the ``true'' number of clusters, 
but rather for the maximum number of clusters which can be resolved from a finite data set. 
This number equals the true number only if there exists a true number of classes and if the data set is 
also large enough to allow us to resolve them.

\section{Rate distortion theory and the Information Bottleneck Method}
If data $x \in X$ are chosen from a probability distribution $P(x)$, then a complete
description of a single data point requires an average code length
equal to the entropy of the distribution, $S(x) = -\sum_x P(x)\log_2 \left[P(x)\right]$
bits.  On the other hand, if we assign points to clusters
$c \in \{ 1, 2, \cdots, N_c \}$, then we need at most $\log_2 (N_c)$ bits.
For  $N_c << |X|$ we have $\log_2 (N_c) << S(x)$, and our intuition is that
many  problems will allow substantial compression at little cost if we
assign  each  $x$ to a cluster $c$ and approximate $x$ by a representative
$x_c$.
\paragraph{Rate distortion theory} formalizes the cost of approximating the signal $x$ by $x_c$
as the expected value of some distortion function,
$d(x,x_c)$  (Shannon 1948).  This distortion measure can, but need not be
a metric. Lossy compression is achieved by assigning the data to clusters such that
the mutual information
\begin{equation}I(c;x) = \sum_{x c} P(c|x) P(x) \log_2 \left[ \frac{P(c|x)}{P(c)} \right] \label{MI}\end{equation}
is minimized. The minimization is constrained by fixing the expected distortion
\begin{equation}\langle d(x,x_c)\rangle = \sum_{x c} P(c|x) P(x) d(x,x_c).\end{equation}
This leads to the variational
problem
\begin{equation}
\min_{P(c|x)} \left[ \langle d(x,x_c)\rangle + T I(c;x)\right].
\end{equation}
The (formal) solution is a Boltzmann distribution\footnote{$T' = T/\ln(2)$, because the information is measured in bits in Eq. (\ref{MI}). $P(c)$ is calculated as $P(c) = \sum_x P(c|x)P(x)$.}
\begin{equation}
P(c|x) =  \frac{P(c)}{Z(x;T)} \exp\left[ -\frac{1}{T'}d(x,x_c)\right],
\label{boltz1}
\end{equation}
with the distortion playing the role of energy, and the normalization
\begin{equation} Z(x,T) = \sum_c P(c) \exp\left[-\frac{1}{T'} d(x,x_c) \right] \end{equation}
playing the role of a partition function (Rose, Gurewitz and Fox, 1990). 
The representatives, $x_c$, often simply called cluster centers, are
determined by the condition
that all of the ``forces'' within each cluster balance for a test point
located at the cluster center,\footnote{This condition is not independent of the original variational problem. Optimizing the objective function with respect to $x_c$, we find: $\frac{\partial}{\partial x_c}\left[ \langle d(x,x_c)\rangle + T I(c;x)\right] = 0 \Leftrightarrow
 \frac{\partial}{\partial x_c}\langle d(x,x_c)\rangle = p(c) \sum_x P(x|c) \frac{\partial}{\partial x_c} d(x,x_c) = 0; \; \forall c \Rightarrow$ Eq. (\ref{centers})}
\begin{equation}
 \sum_x P(x|c) \frac{\partial}{\partial x_c} d(x,x_c) = 0 .\label{centers}
\end{equation}
Recall that if the distortion measure is the squared distance,
$d(x,x_c) = (x-x_c)^2$, then Eq. (\ref{centers}) becomes $x_c = \sum_x x P(x|c)$; the
cluster center is in fact the center of mass of the points which are assigned to the cluster.\\
\\ The Lagrange parameter $T$ regulates the trade--off between the detail we
keep and the bit cost we are willing to pay; in analogy with statistical
mechanics, $T$ often is referred to as the temperature (Rose, Gurewitz and Fox, 1990).
$T$ measures the softness of the cluster membership. The deterministic limit
($T \to 0$) is the limit of hard clustering solutions.
As we lower $T$ there are phase transitions among solutions with
different numbers of distinct clusters, and if we follow these transitions we can
trace out a curve of $\langle d \rangle$ vs. $I(c;x)$, both evaluated at
the minimum. This is the rate--distortion curve and is analogous to plotting energy
vs. (negative) entropy with temperature varying parametrically along the
curve.  Crucially, there is no optimal temperature which provides the
unique best clustering, and thus there is no optimal number of
clusters: more clusters always provide a more detailed description of the
original data and hence allow us to achieve smaller average values of the
distortion $d(x,x_c)$, while the cost of the encoding increases.\\

\paragraph{Information Bottleneck method}
The distortion function {\it implicitly} selects the features that are relevant
for the compression. However, for many problems, we know {\it explicitly} what
it is that we want to keep information about while compressing the data,
but one can not always construct the distortion function that selects
for these relevant features. 
In the Information Bottleneck method (Tishby, Pereira and Bialek, 1999) the relevant
information in the data is defined as information about another variable,
$v \in V$. Both $x$ and $v$ are random variables and we assume that we know
the distribution of co-occurrences, $P(x,v)$. We wish to compress $x$ into
clusters $c$, such that the relevant information, i.e. the information about $v$,
is maximally preserved. This leads directly to the optimization problem
\begin{equation}
\max_{P(c|x)}\left[I(c; v) - T I(c;x) \right].
\label{ibn_vari}
\end{equation}
One obtains a solution similar to Eq. (\ref{boltz1})
\begin{equation}
P(c|x) = \frac{P(c)}{Z(x,T)} \exp\left[-\frac{1}{T} D_{KL}[P(v|x)\|P(v|c)]\right]
\label{ibn_assign}
\end{equation}
in which the Kullback--Leibler divergence,
\begin{equation}
D_{KL}[P(v|x) \| P(v|c)] =
\sum_v P(v|x) \log_2 \left[ \frac{P(v|x)}{P(v|c)} \right],
\end{equation}
emerges in the place of the distortion function (Tishby, Pereira and Bialek, 1999), providing a notion of
similarity between the distributions $P(v|x)$ and $P(v|c)$, where $P(v|c)$ is given by
\begin{equation} P(v|c) = \frac{1}{P(c)} \sum_x P(v|x) P(c|x) P(x). \label{markov}\end{equation}
When we plot $I(c;v)$ as a function of $I(c;x)$, both evaluated at the optimum, we obtain a
curve similar to the Rate Distortion Curve, the slope of which is given by the trade--off between 
compression and preservation of relevant information:
\begin{equation}
\frac{ \delta I(c;v)}{ \delta I(c;x) } = T.
\label{ibn_slope}
\end{equation}
\section{Finite sample effects}
The formulation above assumes that we know the probability distribution
underlying the data, but in practice we have access only to a finite number
of samples, and so there are errors in
the estimation of the distribution. These random errors produce a
systematic error in the computation of the cost function. The idea here is
to compute the error perturbatively and subtract it from the objective
function. Optimization with respect to the assignment rule is now by definition
insensitive to noise and we should (for the IB) find a value for the trade--off
parameter $T^*$ at which the relevant information is kept maximally.\\
\\ The compression problem expressed in Eq. (\ref{ibn_vari}) gives us the
right answer if we evaluate the functional (\ref{ibn_vari})
at the true distribution $P(x,v)$. But in practice we do not know
$P(x,v)$, instead we have to use an estimate $\hat{P}(x,v)$ based on a finite data set. We use perturbation 
theory to compute the systematic error in the cost function that results from the 
uncertainty in the estimate.\\
\\ We first consider the case that $P(x)$ is known and we have to estimate only
the distribution $P(v|x)$. This is the case in many practical clustering problems,
where $x$ is merely an index to the identity of samples, and hence $P(x)$ is constant,
and the real challenge is to estimate $P(v|x)$. In section \ref{uncertain_px}, we
discuss the error that comes from uncertainty in $P(x)$ and also what happens when we apply this
approach to rate--distortion theory.\\
\\ Viewed as a functional of $P(c|x)$, $I(c;x)$ can have errors arising only from uncertainty in
estimating $P(x)$. Therefore, if $P(x)$ is known, then there is no bias in $I(c;x)$. We assume for
simplicity that $v$ is discrete. Let $N$ be the total number of observations of
$x$ and $v$. For a given $x$, the (average) number of observations of $v$ is then $NP(x)$.
We assume that the estimate $\hat{P}(v|x)$ converges to the true distribution in the limit of large 
data set size $ N \to \infty$. However, for finite $N$, the estimated distribution will differ from 
the true distribution and there is a regime in which $N$ is large enough such that we can approximate 
(compare Treves and Panzeri, 1995)
\begin{equation} \hat{P}(v|x) = P(v|x) + \delta P(v|x),\end{equation} where we assume that $\delta P(v|x)$
is some small perturbation and its average over all possible realizations
of the data is zero
\begin{equation} \left\langle \delta P(v|x) \right\rangle = 0. \label{pert.ave.zero} 
\end{equation}
Taylor expansion of $I^{\rm emp}(c;v):=I(c;v)\vert_{\hat{P}(v|x)}$ around $P(v|x)$ leads to a systematic error $\Delta I(c;v)$:
\begin{equation}
I(c;v)\vert_{\hat{P}(v|x) = P(v|x) + \delta P(v|x)} = I(c;v)|_{P(v|x)} + \Delta I(c;v),
\end{equation}
where the error
\begin{equation}
\Delta I(c;v) = \sum_{n=1}^{\infty} \frac{1}{n!} \sum_v \sum_{x^{(1)}} \dots \sum_{x^{(n)}} \frac{\delta^n I(c;v)}{\prod_{k=1}^{n} \delta P(v|x^{(k)})} \left\langle \prod_{k=1}^{n} \delta P(v|x^{(k)}) \right\rangle    
\end{equation}
with 
\begin{eqnarray}
\frac{\delta^n  I(c;v) }{\prod_{k=1}^{n} \delta P(v|x^{(k)})} = (-1)^{n}(n-2)! \left[ \sum_c \frac{ \prod_{k=1}^{n} P(c|x^{(k)}) }{ (P(c,v))^{n-1} } 
-  \frac{ \prod_{k=1}^{n} P(x^{(k)}) }{ (P(v))^{n-1} } \right]
\end{eqnarray}
is given by
\begin{eqnarray}
\Delta I(c;v) &=& \frac{1}{\ln 2}\sum_{n=2}^{\infty} \frac{(-1)^{n}}{n(n-1)} \sum_{v} \left( \sum_{c} \frac{\langle (\sum_x \delta P(v|x)P(c,x))^n \rangle}{(P(c,v))^{n-1}} \right. \nonumber \\ && \left. - \frac{\langle(\sum_x \delta P(v|x)P(x))^n\rangle}{(P(v))^{n-1}} \right),\label{THEerror_1}   
\end{eqnarray}
Note that the terms with $n=1$ vanish, because of Eq. (\ref{pert.ave.zero}) and that the second term in the sum is constant with respect to $P(c|x)$. \\
\\
Our idea is to subtract this error from the objective function (\ref{ibn_vari}) and to recompute the distribution that maximizes the corrected objective function.
\begin{equation}
\max_{P(c|x)}\left[I^{\rm emp}(c; v) - T I^{\rm emp}(c;x) - \Delta I(c;v) + \mu(x) \sum_c P(c|x) \label{obj_corr}\right].
\end{equation}
The last constraint ensures normalization, and the optimal assignment rule $P(c|x)$ is now given by 
\begin{eqnarray}
P(c|x) &=& \frac{P(c)}{Z(x,T)} \exp\left[-\frac{1}{T} \bigg( D_{KL}[P(v|x) \|P(v|c)] + \sum_v  \sum_{n=2}^{\infty} \frac{(-1)^n}{\ln(2)} \right. \nonumber  \\
&& \left. \times \left[ P(v|x) \frac{\langle (\delta P(v|c))^{n}\rangle}{n (P(v|c))^{n}} - \frac{\langle \delta P(v|x)(\delta P(v|c))^{n-1}\rangle}{(n-1) (P(v|c))^{n-1}}\right] \bigg) \right] \label{full_pcx}
\end{eqnarray}
which has to be solved self consistently together with Eq. (\ref{markov})
and
\begin{equation}
\delta P(v|c) := \sum_x \delta P(v|x) P(x|c)
\end{equation}\\

The error $\Delta I(c;v)$ is calculated in Eq. (\ref{THEerror_1}) as an asymptotic expansion and we are assuming that $N$
is large enough to ensure that $\delta P(v|x)$ is small $\forall v$. Let us thus concentrate on the term of leading order
in $\delta P(v|x)$
which is given by (disregarding the term which does not depend on $P(c|x)$)\footnote{We arrive at Eq. (\ref{lead_error}) by calculating the first leading order term ($n=2$) in the sum of Eq. (\ref{THEerror_1}): 
\[ \left(\Delta I(c;v)\right)^{(2)} = \frac{1}{2\ln(2)} \sum_{vc} \frac{\sum_x \sum_x' P(c,x) P(c,x') \langle \delta P(v|x) \delta P(v|x') \rangle}{P(c,v)},\] 
making use of approximation (\ref{approx}) and summing over $x'$, which leads to 
\[ \left(\Delta I(c;v)\right)^{(2)} \simeq \frac{1}{2\ln(2)N} \sum_{vc} \frac{\sum_x [P(c|x)]^2 P(v|x) P(x)}{P(c|v) P(v)},\] and then substituting $P(v|x) P(x)/ P(v) = P(x|v)$ and $P(c|v) = \sum_x P(c|x)P(x|v)$ (compare Eq. (\ref{markov})).} 
\begin{equation} \left(\Delta I(c;v)\right)^{(2)} = \frac{1}{2\ln(2)N} \sum_{vc} \frac{\sum_x [P(c|x)]^2 P(x|v)}{\sum_x P(c|x) P(x|v)},
\label{lead_error} \end{equation}
where we have made use of Eq. (\ref{markov}) and the approximation (for counting statistics)
\begin{equation} \langle \delta P(v|x)\delta P(v|x') \rangle  \simeq \delta_{x x'} \frac{P(v|x)}{N P(x)}.  \label{approx} \end{equation}\\

Can we say something about the shape of the resulting ``corrected'' optimal information curve by analyzing the leading order 
error term (Eq. \ref{lead_error})? This term is bounded from above by the value it assumes in the {\it deterministic limit} 
($T \to 0$), in which assignments $P(c|x)$ are either 1 or 0 and thus $[P(c|x)]^2 = P(c|x)$,\footnote{Substitution of 
$[P(c|x)]^2 = P(c|x)$ into Eq. (\ref{lead_error}) gives $\left(\Delta I(c;v)\right)^{(2)} = \frac{1}{2\ln(2)N} \sum_{vc} \frac{\sum_x P(c|x) P(x|v)}{\sum_x P(c|x) P(x|v)} = \frac{1}{2\ln(2)N} \sum_{vc} = \frac{1}{2\ln(2)N} K_v N_c$.}
\begin{equation} \left(\Delta I(c;v)\right)^{(2)}_{T\to 0} = \frac{1}{2\ln(2)}\frac{K_v}{N} N_c. \label{error.ub}\end{equation}
$K_v$ is the number of bins we have used to obtain our
estimate $\hat{P}(v|x)$. Note that if we had adopted a continuous, rather than a discrete, treatment, then the volume of the (finite)
$V$-space would arise instead of $K_v$.\footnote{For choosing the number of bins, $K_v$, as a function of the data set size $N$, we refer to the large body of literature on this problem, for example Hall and Hannan (1988).}
If one does not make the approximation (\ref{approx}) and, in addition, also keeps the terms constant in $P(c|x)$, then one
obtains for the upper bound ($T \to 0$ limit) of the $n=2$ term from expression (\ref{THEerror_1}):
\begin{equation}
\frac{1}{2\ln(2)}\frac{K_v-1}{N} (N_c - 1)
\end{equation}
which is the leading correction to the bias as derived in (Treves and Panzeri, 1995; Eq. (2.11); term $C_1$). Similar to what these authors found when they computed higher order corrections we also found in numerical experiments that the leading term is a surprisingly good estimate of the total bias and we therefore feel confident to approximate the error by (\ref{lead_error}), although we can not guarantee convergence of the series in (\ref{THEerror_1}).\footnote{ For counting statistics (binomial distribution) we have 
\begin{eqnarray}
\langle (\delta P(v|x))^n \rangle &=& \frac{1}{N^{n-1}}(P(v|x))^n \sum_{k=0}^n (-1)^{(n-k)} \frac{n!}{k!(n-k)!} \frac{N^k}{(P(v|x))^k} \nonumber \\
&&  \times \sum_{\{l_1...l_k\}} \frac{k!N!}{(N-l)!} \left( 1-p \right)^{(N-l)} \prod_{q=1}^k \frac{1}{l_q!} \left(\frac{p}{q!}\right)^{l_q}\nonumber
\end{eqnarray} where $l = \sum_{q=1}^k l_q$, the $l_q$ are positive integers, and the sum $\sum_{\{l_1...l_k\}}$ runs over all partitions of $k$, i.e., values of $l_1$, ...,$l_k$  such that $\sum_{q=1}^k q l_q = k$. There is a growing number of contributions to the sum at each order $n$, some of which can be larger than the smallest terms of the expression at order $n-1$. If there are enough measurements such that $NP(x)$ is large, the binomial distribution approaches a normal distribution with $\langle (\delta P(v|x))^{2n} \rangle = (2n-1)!! \frac{1}{N^{n}P(x)^{n}} \left( P(v|x) - P(v|x)^2 \right)^{n}$, and $\langle (\delta P(v|x))^{2n-1} \rangle = 0$ ($n=1,2,...$).
Substituting this into (\ref{THEerror_1}), and considering only terms with $x^{(1)} = x^{(2)} = ... = x^{(n)}$, we get
$\frac{1}{\ln 2} \sum_{xvc} P(c,v) \sum_{k=1}^{\infty} \frac{(2k-1)!!}{2k(2k-1)}
\left[ \frac{1}{N} \frac{P(x)}{P(v)} \frac{(P(c|x))^{2} }{(P(c|v))^{2}} \left(P(v|x)-P(v|x)^2 \right) \right]^k$, which is not guaranteed to converge.}

The lower bound of the leading order error (Eq. \ref{lead_error}) is given by \footnote{\mbox{Proof: $\sum_{xvc} \frac{[P(c|x)]^2 P(x|v)}{P(c|v)} = \sum_{xc} P(x,c) \frac{P(c|x)}{P(c)} \sum_v \frac{P(v|x)}{P(v|c)}$} $>  \sum_{xc} P(x,c) \frac{P(c|x)}{P(c)} =  \sum_{xc} P(x,c) 2^{\log_2[\frac{P(c|x)}{P(c)}]} \geq 2^{I(c;x)}$}
\begin{equation} \frac{1}{2\ln(2)}\frac{1}{N} 2^{I(c;x)}, \label{error.lb} \end{equation}
and hence the ``corrected information curve'', which we define as
\begin{equation} I^{\rm corr}(c;v) := I^{\rm emp}(c;v) - \Delta I(c;v), \end{equation}
is (to leading order) bounded from above by
\begin{equation} I^{\rm corr}_{UB}(c;v) = I^{\rm emp}(c;v) - \frac{1}{2\ln(2)}\frac{1}{N} 2^{I(c;x)}. \end{equation}
The slope of this upper bound is $T - 2^{I(c;x)}/2N$
(using Eq. (\ref{ibn_slope})), and there is a maximum at
\begin{equation} T^*_{UB} = \frac{1}{2N}2^{I(c;x)}.\label{TstarUB}\end{equation}
\\
\begin{figure}
\center
\includegraphics[width = 14cm]{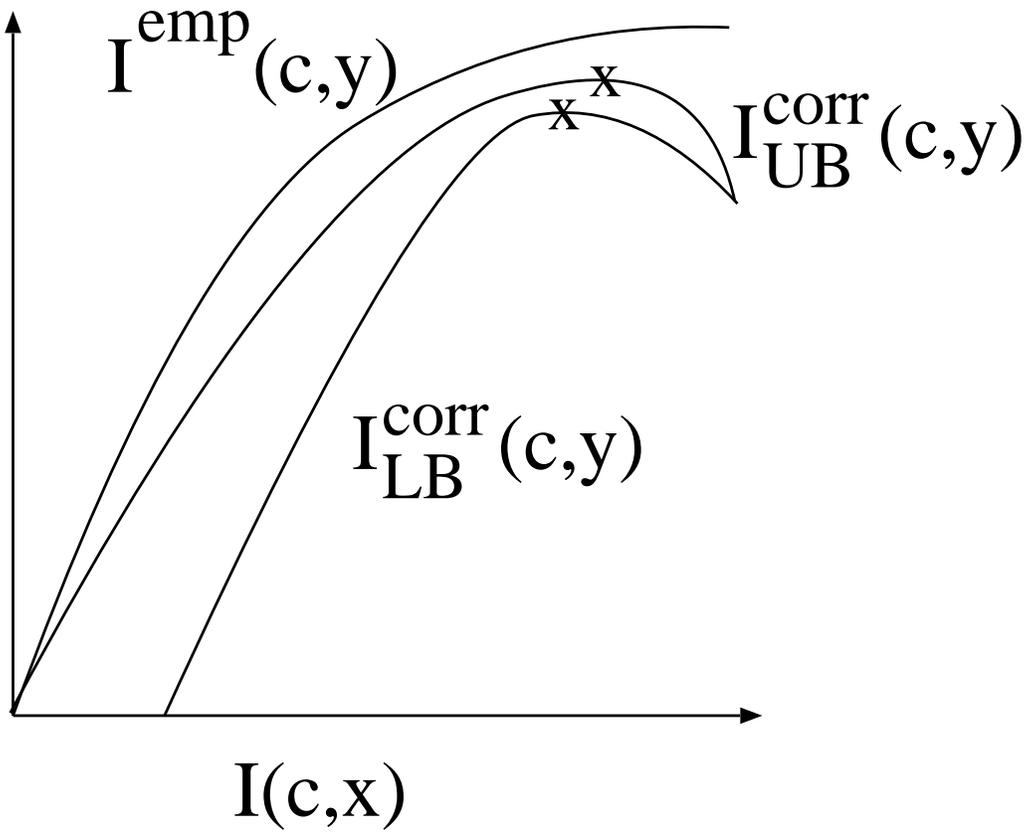}
\caption[]{Sketch of the lower and upper bound on the corrected information curve which both have a maximum under some conditions (see Eqs. (\ref{TstarUB}) and (\ref{TstarLB})), indicated by $x$-es, compared to the empirical information curve which is monotonically increasing.}
\label{sketchfig}
\end{figure}
{\bf If the hard clustering solution assigns equal numbers of data to
each cluster}, then the upper
bound on the error, Eq. (\ref{error.ub}), can be rewritten as
\begin{equation} \frac{1}{2\ln(2)}\frac{K_v}{N} 2^{I(c;x)}.\end{equation}
and therefore, the lower bound on the information curve,
\begin{equation} I^{\rm corr}_{LB}(c;v) = I^{\rm emp}(c;v) - \frac{1}{2\ln(2)}\frac{K_v}{N} 2^{I(c;x)}, \end{equation}
has a maximum at 
\begin{equation} T^*_{LB} = \frac{K_v}{2N}2^{I(c;x)}.\label{TstarLB}\end{equation}
Since both upper and lower bound coincide at the endpoint of the curve, where $T \to 0$ (see sketch in Fig. \ref{sketchfig}),
the actual corrected information curve must have a maximum at 
\begin{equation} T^* = \frac{\gamma}{2N}2^{I(c;x)}.\end{equation}
where $1 < \gamma < K_v$. \\
\\ {\bf In general}, for deterministic assignments, the information we gain by
adding another cluster saturates for large $N_c$, and it is reasonable to assume that this information grows sub-linearly in the number of clusters.
That means that the lower bound on $I^{\rm corr}(c;v)$ has a maximum (or at least a plateau). This
ensures us that $I^{\rm corr}(c;v)$ must have a maximum (or plateau), and hence that an optimal temperature exists.\\
\\ In the context of the IB, asking for the number of clusters that are consistent with the uncertainty
in our estimation of $P(v|x)$ makes sense only for deterministic assignments.
From the above discussion, we know the leading order error term in the
deterministic limit, and we define the ``corrected relevant information'' in the 
limit $T \to 0$ as:\footnote{This quantity is not strictly an information anymore, thus the quotes.}
\begin{equation} I^{\rm corr}_{T \to 0}(c;v) = I^{\rm emp}_{T \to 0}(c;v) - \frac{K_v}{2\ln(2)N} N_c, \label{corr_info}\end{equation}
where $I^{\rm emp}_{T \to 0}(c;v)$ is calculated by fixing the number of clusters and cooling the temperature down to
obtain a hard clustering solution.
While $I^{\rm emp}_{T \to 0}(c;v)$ increases monotonically with $N_c$, we expect
$I^{\rm corr}_{T \to 0}(c;v)$ to have a maximum (or at least a plateau)
at $N_c^*$, as we have argued above.
$N_c^*$, is then the optimal number of clusters in the sense that using more clusters, we would not 
capture more meaningful structure (or in other words would ``overfit''
the data), and although in principle we could always use fewer
clusters, this comes at the cost of keeping less relevant information $I(c;v)$.

\section{Numerical results}
\begin{figure}
\includegraphics[width = 14cm]{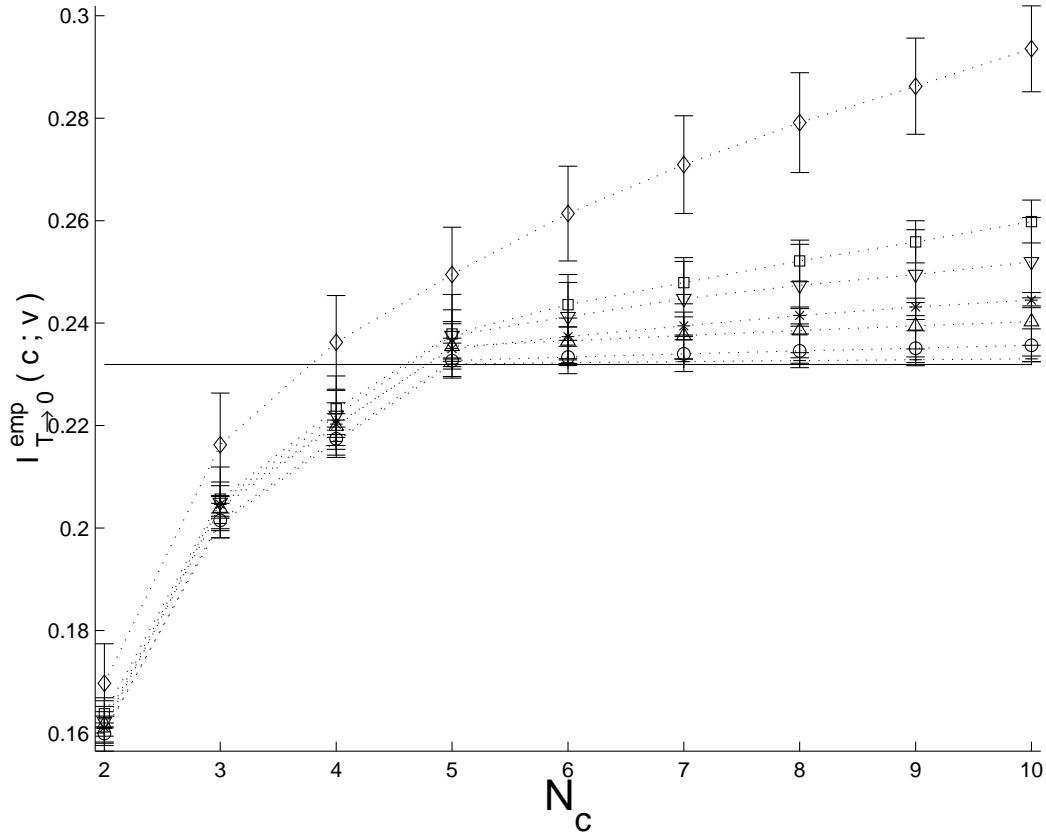}
\caption[]{Result of clustering synthetic data with
$P(v|x) = {\cal N}(0,\alpha(x))$; 5 possible values for $\alpha$.
Displayed is the relevant information kept in the compression, computed from the empirical distribution, $I^{\rm emp}_{T \to 0}(c;v)$, which increases monotonically as a function of the number of clusters. Each curve is computed as the mean of 5 different realizations of the data with error bars of $\pm$ 1 standard deviation. $N_v/K_v$ equals 1 (diamonds), 2 (squares), 3 (down pointing triangles), 5 (stars), 10 (up pointing triangles), 15 (circles) and 50 (crosses). $N_x = 50$ and $K_v = 100$ for all curves. The line is drawn at the value of the information $I(x;v)$, estimated from $10^6$ data points.}
\label{res_5clus_1}
\end{figure}
\begin{figure}
\includegraphics[width = 14cm]{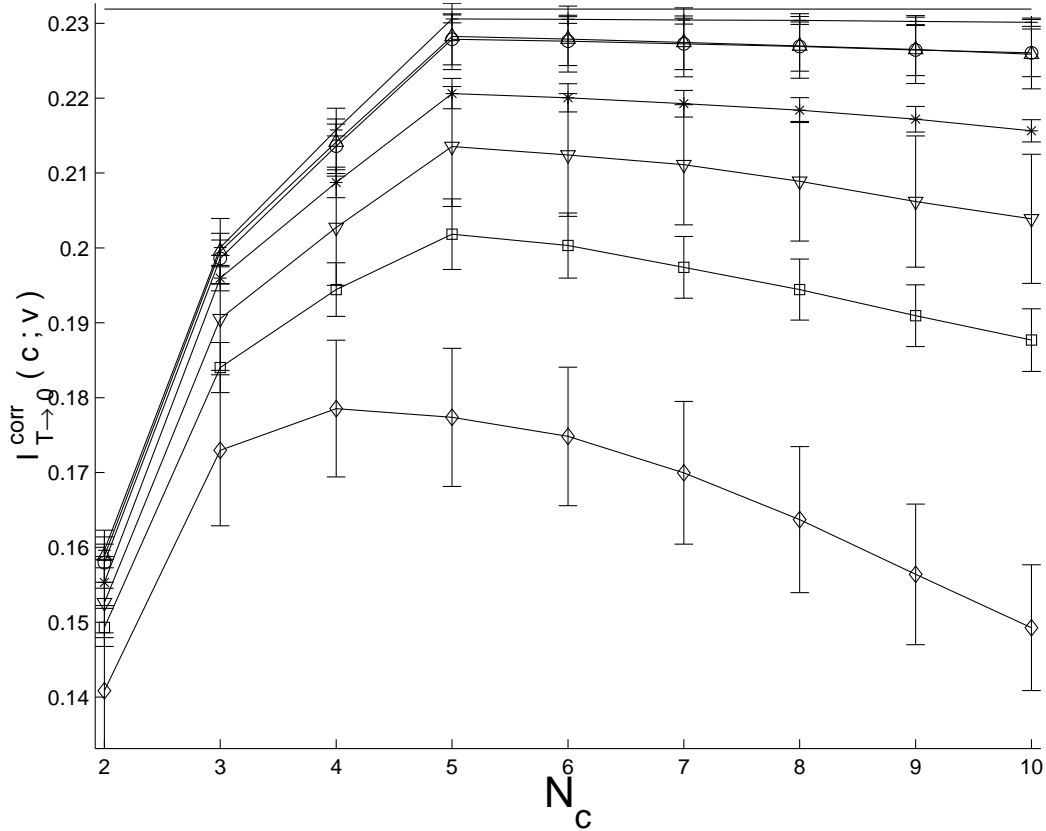}
\caption[]{Result of clustering synthetic data with
$P(v|x) = {\cal N}(0,\alpha(x))$; 5 possible values for $\alpha$.
Displayed is the ``corrected relevant information information'' in the hard clustering limit, $I^{\rm corr}_{T \to 0}(c;v)$, (see Eq. \ref{corr_info}) as a function of the number of clusters. Each curve is computed as the mean of 5 different realizations of the data with error bars of $\pm$ 1 standard deviation. $N_v/K_v$ equals 1 (diamonds), 2 (squares), 3 (down pointing triangles), 5 (stars), 10 (up pointing triangles), 15 (circles) and 50 (crosses). All individual curves (not just the means) peak at $N^*_c = 5$, except for those with $N_v/K_v = 1$, which peak at  $N^*_c = 4$. $N_x = 50$ and $K_v = 100$ for all curves. The line is drawn at the value of the information $I(x;v)$, estimated from $10^6$ data points.}
\label{res_5clus_2}
\end{figure}
\begin{figure}
\includegraphics[width = 14cm]{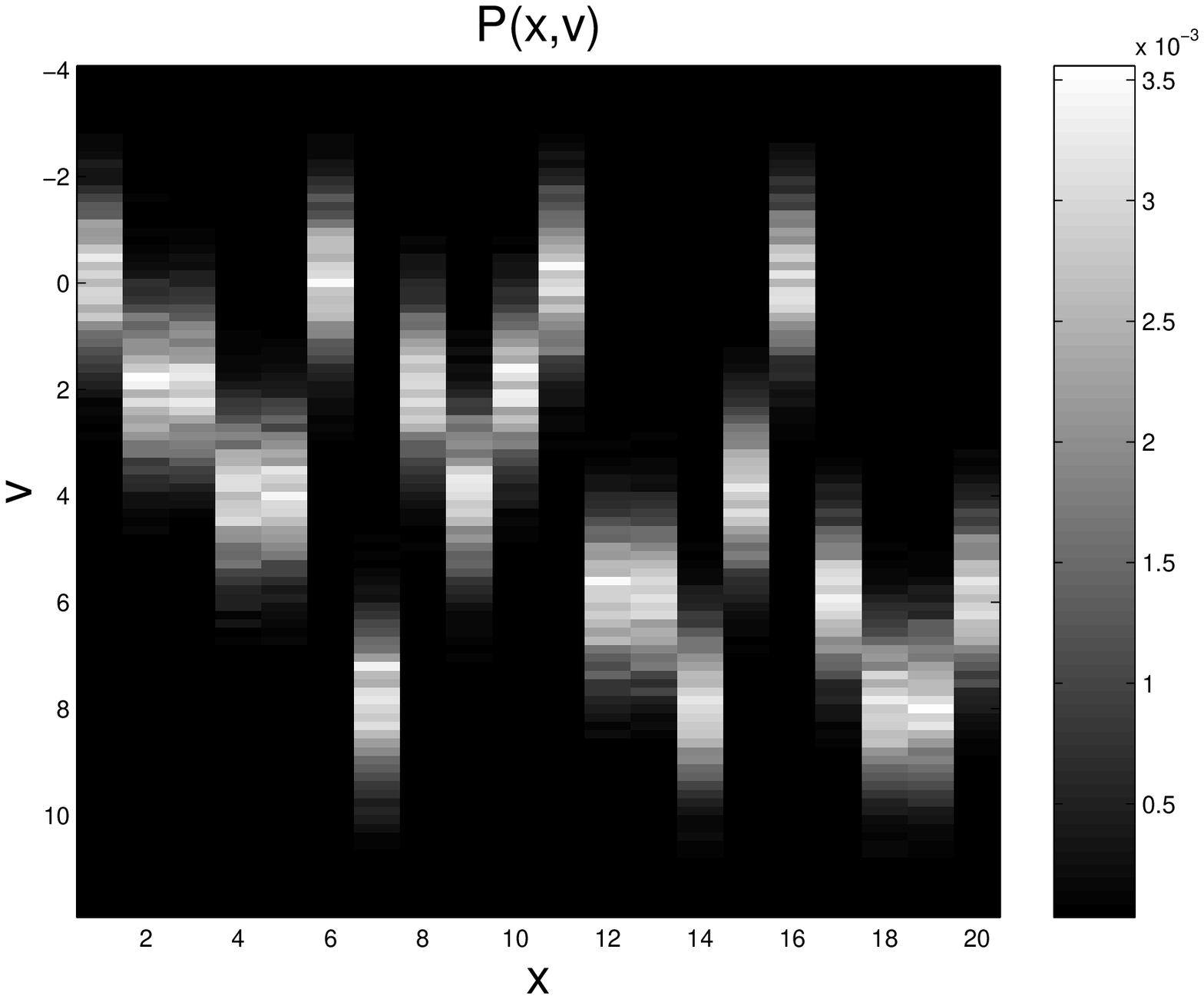}
\caption[]{A trivial examples of those data sets on which we found the correct number of clusters (results are summarized in Fig. \ref{PLANEgauss_space_N_2_5_10}). Here, $P(v|x) = {\cal N}(\alpha(x), 1)$ with 5 different values for $\alpha$, spaced $d\alpha = 2$ apart. $K_v = 100$, $N_x = 20$, $N_v/K_v = 20$.}
\label{DATAeasy}
\end{figure}
\begin{figure}
\includegraphics[width = 14cm]{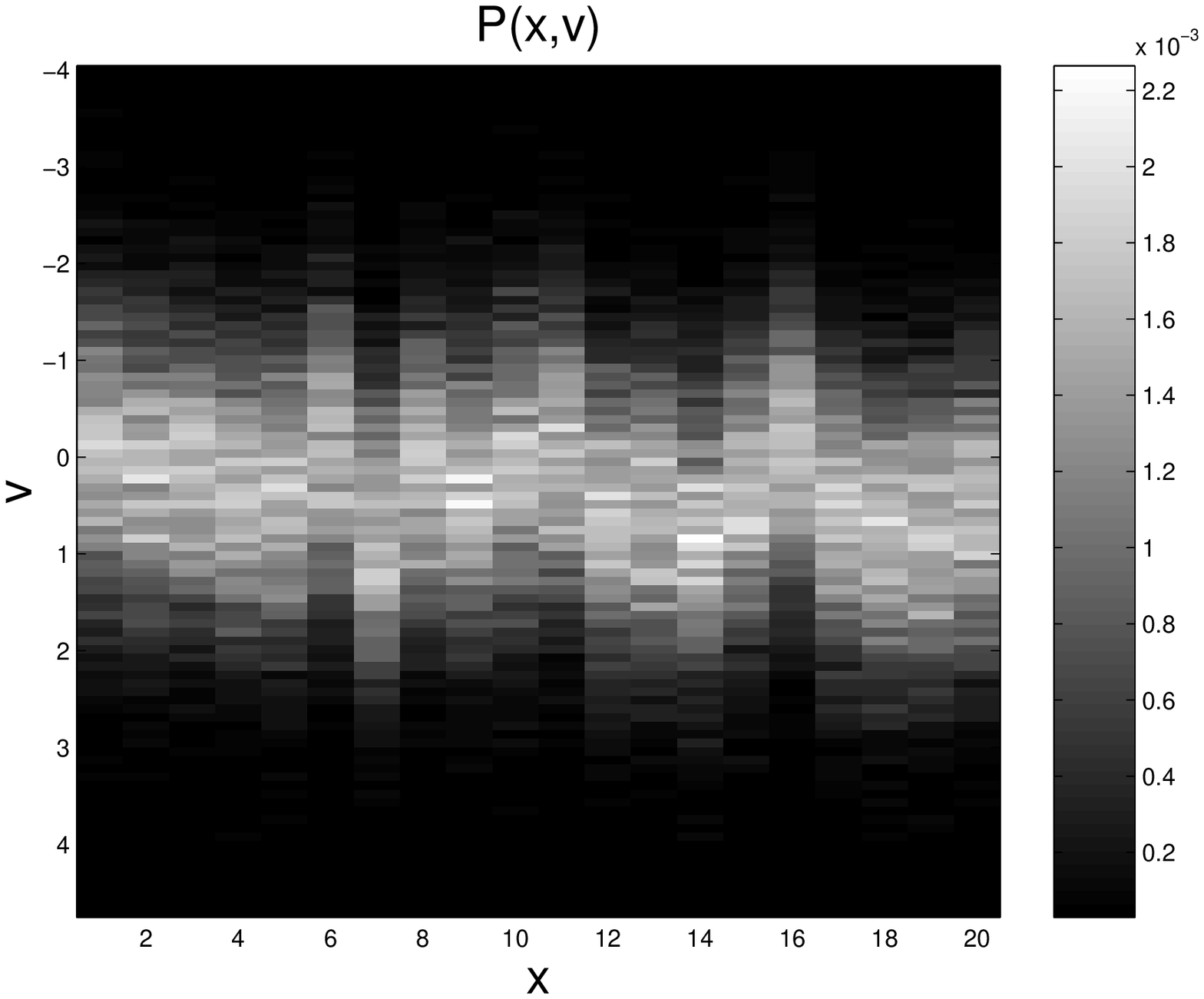}
\caption[]{One of the difficult examples of those data sets on which we found the correct number of clusters (results are summarized in Fig. \ref{PLANEgauss_space_N_2_5_10}). Here, $P(v|x) = {\cal N}(\alpha(x), 1)$ with 5 different values for $\alpha$, spaced $d\alpha = 0.2$ apart. $K_v = 100$, $N_x = 20$, $N_v/K_v = 20$.}
\label{DATA}
\end{figure}
\begin{figure}
\includegraphics[width = 14cm]{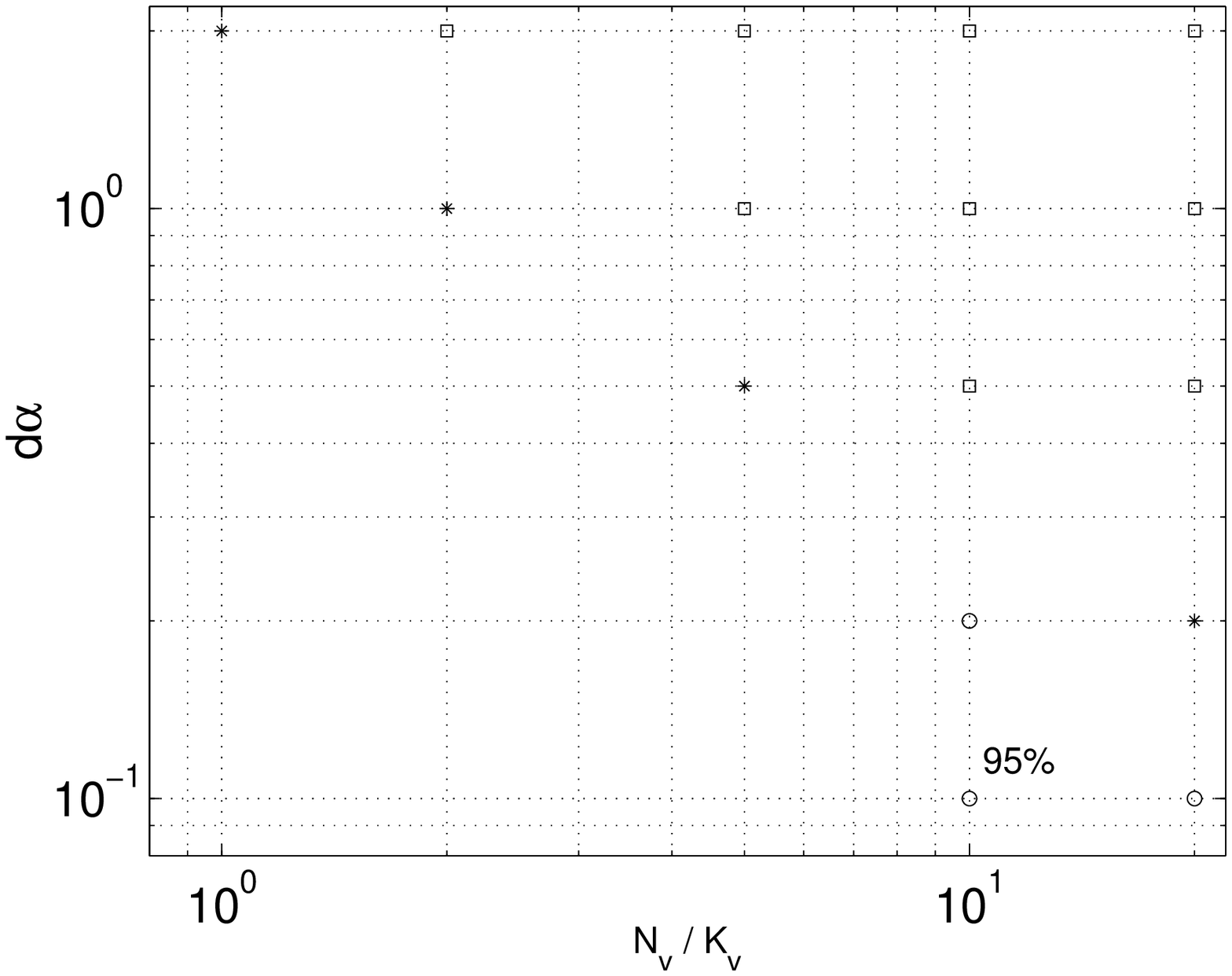}
\caption[]{Result of finding the correct number of clusters with our method
for a synthetic data set of size $N = N_x N_v$ ($N_x = 20$) with
$P(v|x) = {\cal N}(\alpha(x), 1)$ and with either 2, 5 or 10 possible values
for $\alpha$, spaced $d\alpha$ apart. We indicate values of $d\alpha$ and
the resolution $N_v/K_v$ ($K_v = 100$) at which the correct number of
clusters is found: for 2, 5 and 10 clusters (squares); only for 2 and 5
clusters (stars); only for 2 clusters (circles). The classification error (on the training data)
is 0 for all points except for the one that is labeled with $95\%$ correct.}
\label{PLANEgauss_space_N_2_5_10}
\end{figure}

\subsection{Simple synthetic test data}
We test our method for finding $N^*_c$ on data that we understand well and
where we know what the answer {\it should} be. We thus created synthetic data drawn from normal distributions with zero mean and 
5 different variances (for Figs. \ref{res_5clus_1} and \ref{res_5clus_2}).\footnote{$P(x) = 1 / N_x$
and $P(v|x) = {\cal N}(0,\alpha(x))$ where $\alpha(x) \in A$, and $|A| = 5$,
with $P(\alpha) = 1 / 5$; and $N_x = 50$. $N_x$ is the number of ``objects'' we are clustering.}
We emphasize that we chose an example with Gaussian distributions {\it not} because any of our analysis makes use of Gaussian assumptions, but rather because in the Gaussian case we have a clear intuition about the similarity of different distributions and hence about the difficulty of the clustering task. This will become important later, when we make the discrimination task harder
(see Fig. \ref{PLANEgauss_space_N_2_5_10}). We use $K_v = 100$ bins to estimate $\hat{P}(v|x)$.
In Figs. \ref{res_5clus_1} and \ref{res_5clus_2}, we compare how $I_{T \to 0}^{\rm emp}(c;v)$ and
$I^{\rm corr}_{T \to 0}(c;v)$ behave as a function of the
number of clusters. The number of observations of $v$, given $x$, is
$N_v = N/N_x$. For a large range of ``resolutions'' (average number of observations per bin), $N_v/K_v$,
$I^{\rm corr}_{T \to 0}(c;v)$ has a maximum at $N^*_c = 5$ (the true number of clusters). When we have too little data ($N_v/K_v = 1$), 
we can resolve only 4 clusters. The curves in Figs. \ref{res_5clus_1} and \ref{res_5clus_2} are calculated as the mean of
the curves obtained from 5 different realizations of the data,\footnote{Each time we compute $I_{T \to 0}^{\rm emp}(c;v)$, we start at 100 different, randomly chosen initial conditions to increase the probability of finding the global maximum of the objective functional.} and the error bars are $\pm$ 1 standard deviation. The optimal number of clusters $N_c^*$, as determined by our method, is always the same, for each of these individual curves, so that there are no error 
bars on the optimal number of clusters as a function of the data set size. As $N_v/K_v$ becomes very large, $I^{\rm corr}_{T \to 0}(c;v)$
approaches $I_{T \to 0}^{\rm emp}(c;v)$, as expected. \\
\\ The curves in Figs. \ref{res_5clus_1} and \ref{res_5clus_2} differ in the average number of examples
per bin, $N_v/K_v$. The classification problem becomes harder as we see less data.
However, it also becomes harder when the true distributions are more like each
other. To separate the two effects, we create synthetic data drawn from 
Gaussian distributions with unit variance and $N_A$ different, equidistant means $\alpha$,
which are
$d \alpha$ apart.\footnote{$P(v|x) = {\cal N}(\alpha(x), 1)$, $\alpha(x) \in A$, $N_A:=|A|$.} 
$N_A$ is the true number of clusters. This problem becomes intrinsically harder as
$d \alpha$ becomes smaller. Examples are shown in Figs. \ref{DATAeasy} and 
\ref{DATA}. The problem becomes easier as we are allowed to look at more data, 
which corresponds to an increase in $N_v/K_v$.
We are interested in the regime in the space spanned by $N_v/K_v$ and $d \alpha$
in which our method retrieves the correct number of clusters.\\
\\ In Fig. \ref{PLANEgauss_space_N_2_5_10}, points mark those values of
$d \alpha$ and $N_v/K_v$ (evaluated on the shown grid) at which we find
the true number of clusters. The different shapes of the points summarize
results for 2, 5 and 10 clusters. A missing point on the grid indicates a
value of $d \alpha$ and $N_v/K_v$ at which we did not find the correct number
of clusters. All these missing points lie in a regime which is characterized
by a strong overlap of the true distributions combined with scarce data. In
that regime, our method always tells us that we can resolve {\it fewer} clusters than the true number of
clusters. For small sample sizes, the correct number of clusters is resolved
only if the clusters are well separated, but as we accumulate more data, we
can recover the correct number of classes for more and more overlapping
clusters. To illustrate the performance of the method, we show in Fig.
\ref{DATA} the distribution $P(x,v)$ in which $\alpha(x)$ has 5 different
values which occur with equal probability, $P(\alpha(x)) = 1/5$ and which 
differ by $d\alpha = 0.2$. For this separation, our method still retrieves 
5 as the optimal number of clusters when we have $N_v=2000$ observations per example $x$.
\footnote{We used $K_v=100$ bins to estimate $P(v|x)$.The 
distribution of examples is simply $P(x) = 1/N_x$ with $N_x = 20$.} \\
\\ Our method detects when only one cluster is present, a case in which many
methods fail (Gordon, 1999). We verified this for data drawn from one
Gaussian distribution and for data drawn from the uniform distribution.

\begin{figure}
\includegraphics[width = 14cm]{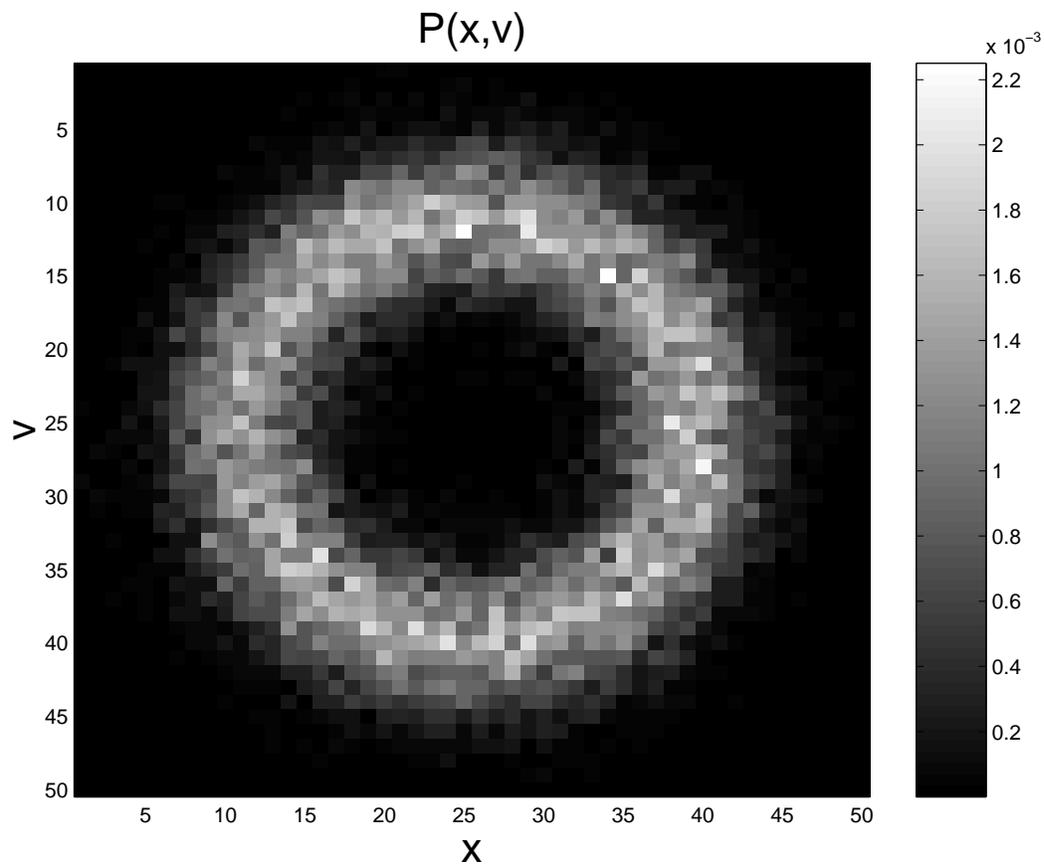}
\caption[]{20000 data points drawn from a radial distribution, according to $P(r) = {\cal N}(1,0.2)$, with $x=r cos(\phi)$, $v=r sin(\phi)$, $P(\phi)= 1/2\pi$. Displayed is the estimated probability distribution (normalized histogram with 50 bins along each axis).}
\label{input_ringdist}
\end{figure}
\begin{figure}
\includegraphics[width = 14cm]{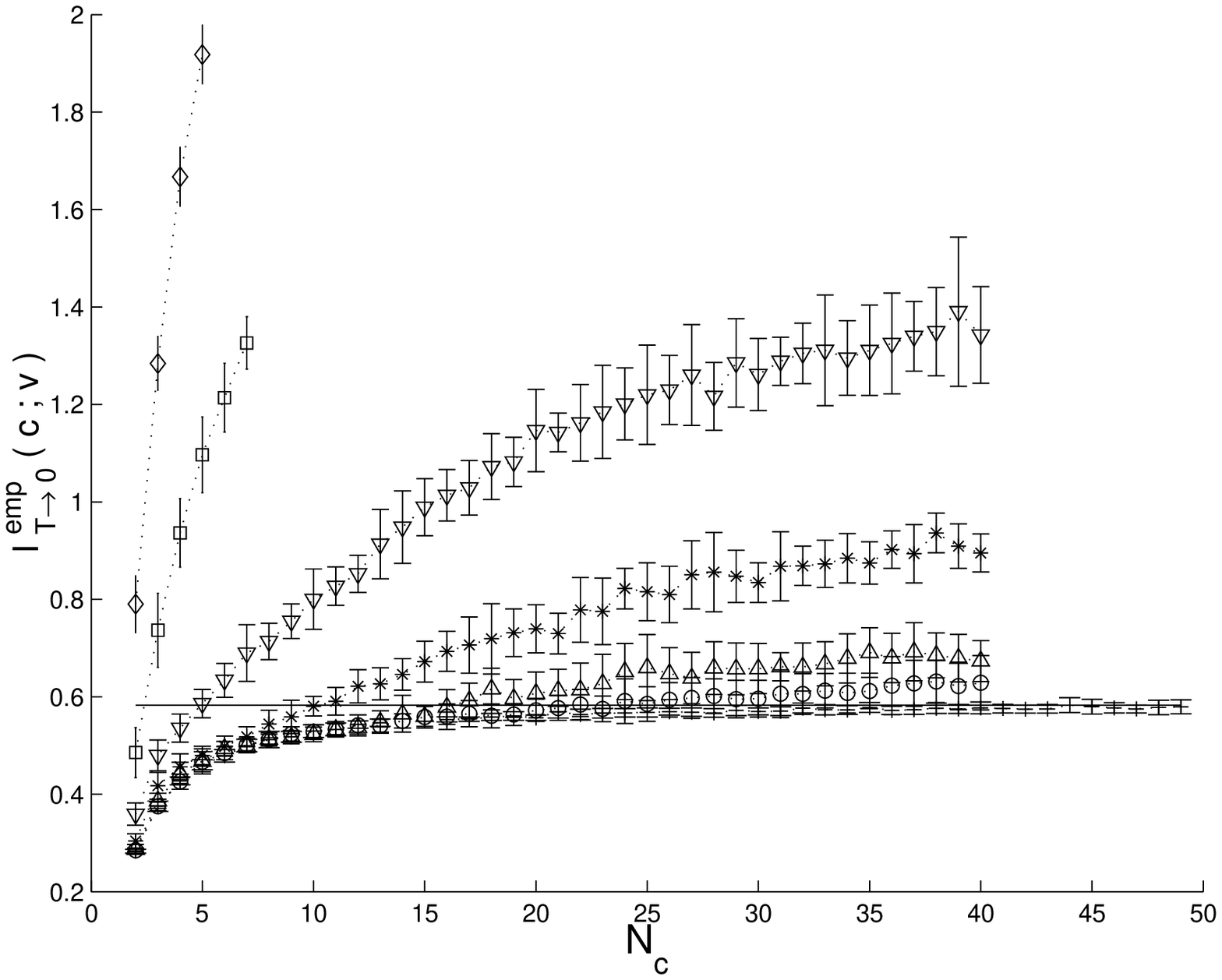}
\caption[]{$I^{\rm emp}_{T \to 0}(c;v)$ as a function of the number of clusters, averaged over 7 different realizations of the data. Error bars are $\pm$ 1 standard deviation. The information $I(x;v)$ calculated from 100000 data points is 0.58 bits (line). Data set size $N$ is: 100 (diamonds), 300 (squares), 1000 (down pointing triangles), 3000 (stars), 10000 (up pointing triangles), 30000 (circles), 100000 (crosses).}
\label{infocurve_emp_ringdist}
\end{figure}
\begin{figure}
\includegraphics[width = 14cm]{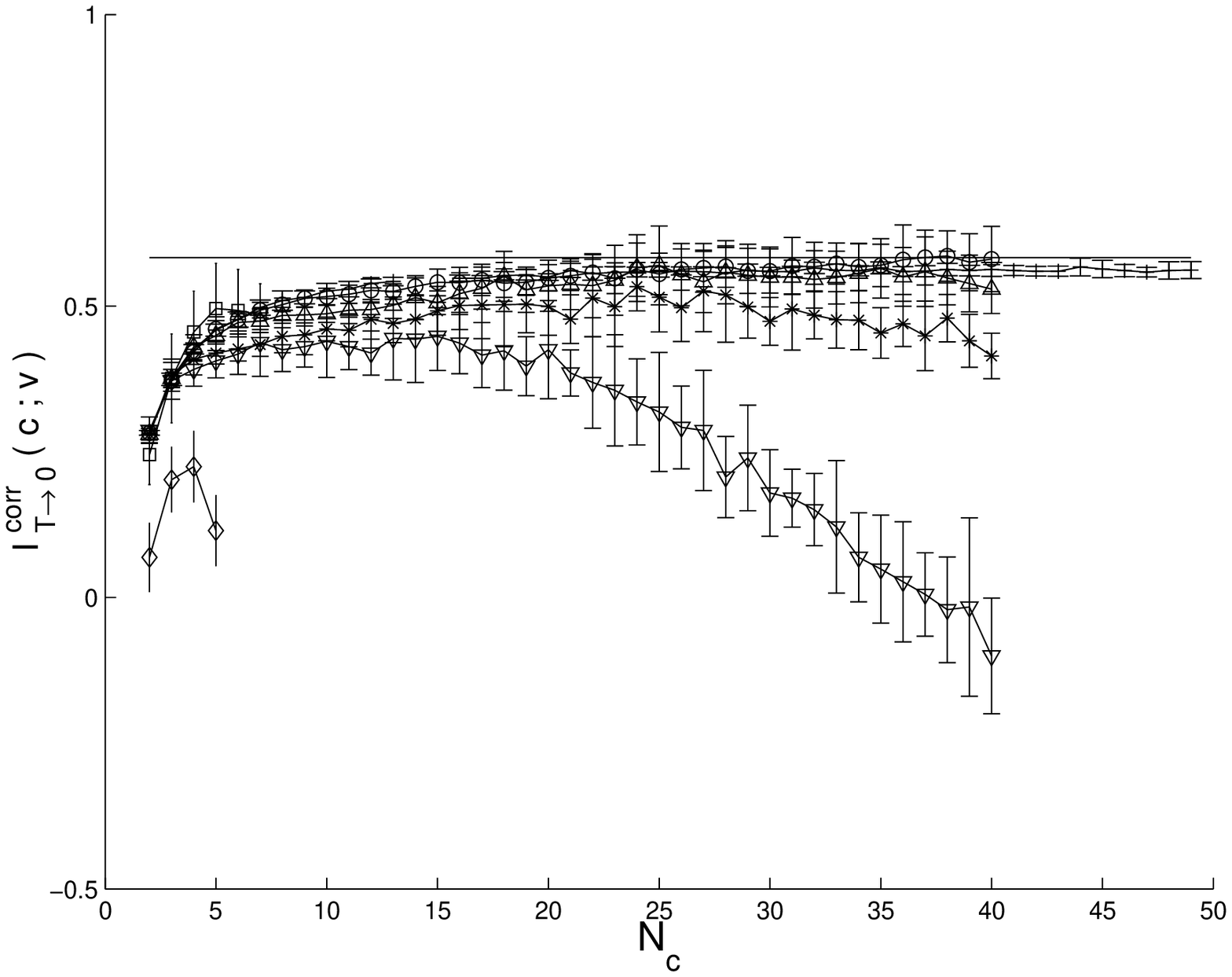}
\caption[]{$I^{\rm corr}_{T \to 0}(c;v)$ as a function of the number of clusters, averaged over 7 different realizations of the data. Error bars are $\pm$ 1 standard deviation. The information $I(x;v)$ calculated from 100000 data points is 0.58 bits (line). Data set size $N$ is: 100 (diamonds), 300 (squares), 1000 (down pointing triangles), 3000 (stars), 10000 (up pointing triangles), 30000 (circles), 100000 (crosses).}
\label{infocurve_corr_ringdist}
\end{figure}
\begin{figure}
\includegraphics[width = 14cm]{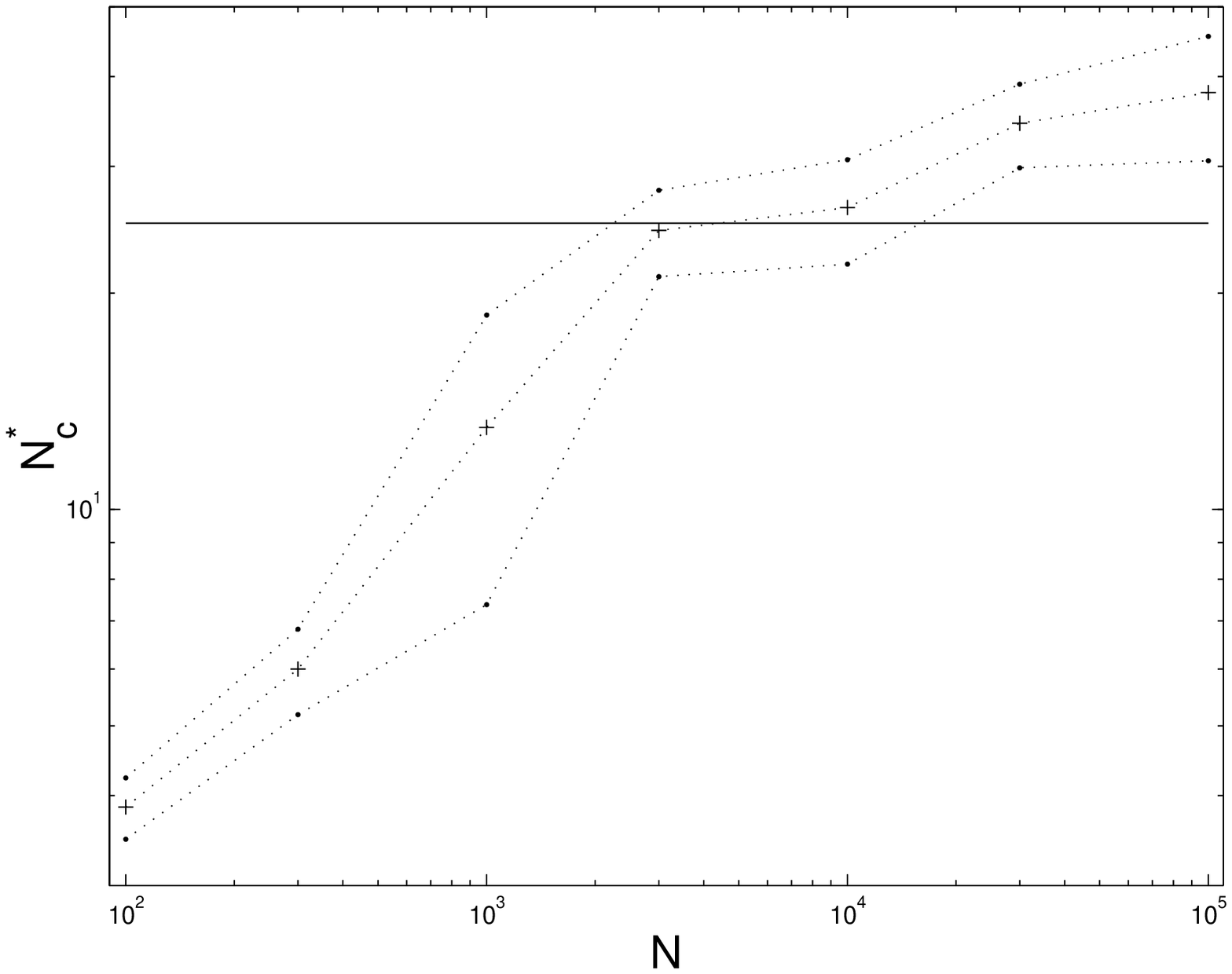}
\caption[]{Optimal number of clusters, $N_c^*$, as found by the suggested method, as a function of the data set size $N$. The middle curve (crosses) represents the average over 7 different realizations of the data, points on the upper/lower curve $\pm$ 1 standard deviation. Line at 25.}
\label{NCcurve_ringdist}
\end{figure}
\subsection{Synthetic test data that explicitly violates mixture model assumptions}
We consider data drawn from a radial normal distribution, according to $P(r) = {\cal N}(1,0.2)$, with $x=r cos(\phi)$, $v=r sin(\phi)$, and $P(\phi)= 1/2\pi$, as shown in Fig. \ref{input_ringdist}. The empirical information curves (Fig. \ref{infocurve_emp_ringdist}) and corrected information curves (Fig. \ref{infocurve_corr_ringdist}) are computed as the mean of 7 different realizations of the data for different sample sizes.\footnote{Each time we compute $I_{T \to 0}^{\rm emp}(c;v)$, we start at 20 different, randomly chosen initial conditions to increase the probability of finding the global maximum of the objective functional. Increasing the number of initial conditions would decrease the error bars at the cost of computational time.} The corrected curves peak at $N_c^*$, which is shown as a function of $N$ in Fig. \ref{NCcurve_ringdist}. For less than a few thousand samples, the optimal number of clusters goes roughly as $N_c^* \propto N^{2/3}$, but there is a saturation around $N_c^* \approx 25$. This number corresponds to half of the number of x-bins (and therefore half of the number of ``objects'' we are trying to cluster) which makes sense given the symmetry of the problem.

\section{Uncertainty in $P(x)$}
\label{uncertain_px}
In the most general case, $x$ can be a continuous variable drawn from an unknown distribution $P(x)$.
We then have to estimate the full distribution $P(x,v)$ and if we want to follow the same
treatment as above, we have to assume that our estimate approximates the true distribution
\begin{equation} \hat{P}(v,x) = P(v,x) + \delta P(v,x),\end{equation} where $\delta P(v,x)$
is some small perturbation and its average over all possible realizations of the data is zero
\begin{equation} \left\langle \delta P(v,x) \right\rangle = 0. \label{pert2.ave.zero} \end{equation}
Now, this estimate induces an error not only in $I^{\rm emp}(c; v)$, but also in $I^{\rm emp}(c;x)$.
Taylor expansion of these two terms gives 
\begin{eqnarray}
\Delta I(c;v) &=& \frac{1}{\ln(2)} \sum_{vc} \sum_{n=2}^{\infty} \frac{(-1)^n}{n(n-1)} \left(\frac{1}{(P(v,c))^{n-1}} - \frac{1}{(P(c))^{n-1}} \right) \nonumber \\ && \times \left\langle \left(\sum_x P(c|x) \delta P(x,v)\right)^n \right\rangle - \Lambda(P(v)) \\
\Lambda(P(v)) &=& \frac{1}{\ln(2)} \sum_{v} \sum_{n=2}^{\infty} \frac{(-1)^n}{n(n-1)} \frac{1}{(P(v))^{n-1}} \left\langle \left( \sum_x \delta P(x,v)\right)^n \right\rangle \\
\Delta I(c;x) &=& - \frac{1}{\ln(2)} \sum_{vc} \sum_{n=2}^{\infty} \frac{(-1)^n}{n(n-1)} \frac{1}{(P(c))^{n-1}} \nonumber\\ && \times \left\langle \left(\sum_x P(c|x) \delta P(x,v)\right)^n \right\rangle 
\end{eqnarray}
This results in a correction to the objective function ($F^{\rm corrected} = F^{\rm emp} -\Delta F$), given by:
\begin{eqnarray}
\Delta F &=& \frac{1}{\ln(2)} \sum_{vc} \sum_{n=2}^{\infty} \frac{(-1)^n}{n(n-1)} \frac{1}{(P(c))^{n-1}} \left( \frac{1}{(P(v|c))^{n-1}} + T-1 \right) \label{errorF} \\ && \times \left\langle \left(\sum_x P(c|x) \delta P(x,v)\right)^n \right\rangle - \Lambda(P(v)), \nonumber
\end{eqnarray}
where $\Lambda(P(v))$ is constant in $P(c|x)$ and therefore not important. At critical temperature $T=1$, the error due to uncertainty in $P(x)$ made in calculating $I^{\rm emp}(c; v)$ cancels that made in computing $I^{\rm emp}(c;x)$. For small $T$, the largest contribution to the error is given by the first term in the sum of Eq. (\ref{errorF}), since $1/(P(v|c))^n \geq 1$, $\forall \{ n,v,c\}$. Therefore, the procedure that we have suggested for finding the optimal number of clusters in the deterministic limit ($T \to 0$) remains unchanged, even if $P(x)$ is unknown. Let us consider, as before, the leading order term of the error (using the approximation in Eq. (\ref{approx}))
 \begin{equation}
\left(\Delta F\right)^{(2)} = \frac{1}{2 N \ln(2)} \sum_{cv} \frac{1}{p(c)} \left( \frac{1}{p(v|c)} + T - 1 \right) \sum_x (P(c|x))^2 P(x,v).
\end{equation}
In the $T \to 0$ limit this term becomes $N_c (K_v -1)/2N\ln(2)$, and we find, 
\begin{equation}
I_{T \to 0}^{\rm corr}(c;v) = I_{T \to 0}^{\rm emp}(c;v) - \frac{K_v - 1}{2\ln(2)N} N_c,
\end{equation}
which is insignificantly different from Eq. (\ref{corr_info}) in the regime $K_v >> 1$.\\
\\
Only for very large temperatures $T >> 1$ (i.e. at the onset of the annealing process) could the error that results from uncertainty in $P(x)$ make a significant difference. \\
\\ The corrected objective function is now given by 
\begin{equation} 
F^{\rm corr} = I^{\rm emp}(c; v) -T I^{\rm emp}(c;x) -\Delta F - \mu(x)\sum_c P(c|x),
\end{equation} and the optimal assignment rule is given by
\begin{eqnarray}
P(c|x) &=& \frac{P(c)}{Z(x,T)} \exp \bigg[-\frac{1}{T} \bigg( D_{KL}\left[P(v|x) \|P(v|c)\right] \nonumber \\
&&+ \frac{1}{\ln(2) P(x)} \sum_v \sum_{n=2}^{\infty} (-1)^n  \bigg[ \frac{1}{n (P(c))^{n}} \left( \frac{1}{(P(v|c))^n} +T-1 \right) \nonumber \\ && \times \left\langle (\delta P(v,c))^{n} \right\rangle - \frac{1}{(n-1) (P(c))^{n-1}} \left( \frac{1}{(P(v|c))^{n-1}} +T-1\right) \nonumber \\ && \times \left\langle \delta P(x,v)(\delta P(v,c))^{n-1} \right\rangle \bigg] \bigg) \bigg],  
\end{eqnarray}
which has to be solved self consistently together with Eq. (\ref{markov})
and
\begin{equation}
\delta P(v,c) := \sum_x \delta P(v,x) P(c|x)
\end{equation}

\paragraph{Rate--distortion theory.}
Let us assume that we estimate the distribution $P(x)$ by 
$\hat{P}(x) = P(x) + \delta P(x)$, with $\langle \delta P(x)\rangle = 0$, as before. While there is no 
systematic error in the computation of $\langle d \rangle$, this uncertainty in $P(x)$ does produce 
a systematic {\em under} estimation of the information cost $I(c;x)$:
\begin{eqnarray}
\Delta I(c;x) = - \frac{1}{\ln(2)} \sum_{n=2}^{\infty} \frac{(-1)^n}{n(n-1)} 
\sum_{c} \frac{\left\langle \left( \sum_x P(c|x) \delta P(x)\right)^n \right\rangle }{(P(c))^{n-1}}
\end{eqnarray}
When we correct the cost functional for this error (with $\lambda = 1/T$),
\begin{equation} F^{\rm corr} := I(c;x) + \lambda \langle d(x,x_c) \rangle - \Delta I(c;x) + \mu (x) \sum_c P(c|x), \end{equation}
we obtain for the optimal assignment rule (with $\lambda' = \lambda \ln(2)$),
\begin{eqnarray}
P(c|x) &= \frac{P(c)}{Z(x,\lambda)} \exp & \left[ -\lambda' d(x,x_c) + \sum_c \sum_{n=2}^{\infty} \left( \frac{ \left\langle \left( \sum_x P(c|x) \delta P(x) \right)^n \right\rangle}{n \left(P(c)\right)^n} \right. \right. \nonumber \\ && - \left. \left. \frac{1}{P(x)} \frac{ \left\langle \delta P(x) \left( \sum_x P(c|x) \delta P(x) \right)^{n-1} \right\rangle}{(n-1) \left(P(c)\right)^{n-1}}\right) \right]. 
\end{eqnarray}
Let us consider the leading order term of the error made in calculating the information cost, 
\begin{equation}
\left(\Delta I(c;x)\right)^{(2)} = - \frac{1}{2 \ln(2) N} \sum_c \frac{\langle \sum_x \left( P(c|x) \delta P(x)\right)^2 \rangle}{P(c)}.
\end{equation}
For counting statistics, we can approximate, as before,  
\begin{equation}
\left(\Delta I(c;x)\right)^{(2)} \approx - \frac{1}{2 \ln(2) N} \sum_c \frac{\sum_x \left(P(c|x)\right)^2 P(x)}{P(c)}.
\end{equation}
The information cost is therefore underestimated by at least $2^{I(c;x)}/2$ bits.\footnote{Using 
$\sum_{xc} P(x,c) \frac{P(c|x)}{P(c)} =  \sum_{xc} P(x,c) 2^{\log_2[\frac{P(c|x)}{P(c)}]} \geq 2^{I(c;x)}$.}
The corrected rate--distortion curve with
\begin{equation}
I^{\rm corr} (c;x) := I(c;x) - \Delta I(c;x)
\end{equation}
is then bounded from below by
\begin{equation}
I^{\rm corr}_{LB} (c;x) = I(c;x) + \frac{1}{2\ln(2)N} 2^{I(c;x)}
\end{equation}
and this bound has a rescaled slope given by
\begin{equation}
\tilde{\lambda} = \lambda' \left( 1- \frac{1}{2N} 2^{I(c;x)} \right)
\end{equation}
but no extremum. Since there is no optimal trade--off, it is not possible to use the same arguments as we have used before to determine an optimal number of clusters in the hard clustering limit. To do this, we have to carry the results obtained from the treatment of the finite sample size effects in the IB over to rate--distortion theory. This is possible, with insights we have gained in Still, Bialek and Bottou (2004) about how to use the IB for data that are given with some measure of distance (or distortion).

\section{Summary}
Clustering, as a form of lossy data compression, is a trade--off between the quality and complexity of representations.  In
general, a data set (or clustering problem) is characterized by the whole structure of this trade--off --- the
rate--distortion curve or the information curve in the IB method --- which quantifies our intuition that some data are
more clusterable than others.  In this sense there is never a single ``best'' clustering of the data, just a family of
solutions evolving as a function of temperature.\\
\\
As we solve the clustering problem at lower temperatures, we find solutions that reveal more and more detailed
structure and hence have more distinct clusters.  If we have only finite data sets, however, we expect that there is an
end to the {\em meaningful} structure that can be resolved --- at some point separating clusters into smaller groups just
corresponds to fitting the sampling noise.  The traditional approach to this issue is to solve the clustering problem
in full, and then to test for significance or validity of the results by some additional statistical criteria.  What we
have presented in this work is, we believe, a new approach:  Because clustering is formulated as an optimization
problem, we can try to take account of the sampling errors and biases directly in the objective functional.  In
particular, for the Information Bottleneck method all terms in the objective functional are mutual informations, and
there is a  large literature on the systematic biases in information estimation.  There is a perturbative regime in
which these biases have a universal form and can be corrected.  Applying these corrections, we find that at fixed
sample size the trade--off between complexity and quality really does have an endpoint beyond which lowering the
temperature or increasing the number of clusters does not resolve more relevant information. We have seen numerically
that in model problems this strategy is sufficient to set the maximum number of resolvable clusters at the correct
value.

\subsection*{Acknowledgments}
We thank N. Tishby for helpful comments on an earlier draft. S. Still wishes to 
thank M. Berciu and L. Bottou for helpful discussions and acknowledges support from
the German Research Foundation (DFG), grant no. Sti197.

\subsection*{References}
\begin{itemize}{}

\item[] V. Balasubramanian, Statistical Inference, Occam's Razor, and Statistical Mechanics on the Space of Probability Distributions. Neural Comp. 9 (1997) 349.

\item[] M. Blatt, S. Wiseman and E. Domany, Superparamagnetic Clustering of Data. Phys. Rev. Lett. 76 (1996) 3251-3254, cond-mat/9702072

\item[] H.-H. Bock, Probability Models and Hypotheses Testing in Partitioning Cluster Analysis. In {\it Clustering and Classification} Eds.: P. Arabie, L.J. Hubert and G. De Soete (1996) World Scientific, pp. 378-453.

\item[] J. M. Buhmann and M. Held, Model Selection in Clustering by Uniform Convergence Bounds. In {\it Adv. Neural Inf. Proc. Sys. (NIPS) 12} Eds.: S. A. Solla, T. K. Leen and K.-R. M\"uller (2000) MIT Press, Cambridge, MA

\item[] M. Eisen, P. T. Spellman, P. O. Brown and D. Botstein, Cluster analysis and display of genome-wide expression patterns. Proc. Nat. Acad. Sci. (USA) 95 (1998) 14863.

\item[] C. Fraley and A. Raftery, Model-based clustering, discriminant analysis, and density estimation. J. Am. Stat. Assoc. 97 (2002) 611.

\item[] A. D. Gordon, {\it Classification}, (1999) Chapmann and Hall/CRC Press, London

\item[] P. Hall and E. J. Hannan, On stochastic complexity and nonparametric density estimation. Biometrika 75 (1988) No. 4, pp.705-714.

\item[] D. Horn and A. Gottlieb, Algorithm for Data Clustering in Pattern Recognition Problems Based on Quantum Mechanics. Phys. Rev. Lett. 88 (2002) 018702, extended version: physics/0107063

\item[] S. Lloyd, Least squares quantization in PCM. Technical Report (1957) Bell Laboratories. Also in: IEEE Trans. Inf. Th., vol. IT-28 (1982) 129.

\item[] J. MacQueen, Some methods for classification and analysis of multivariate observations. In {\it Proc. 5th Berkeley Symp. Math. Statistics and Probability} Eds.: L.M.L Cam and J. Neyman (1967) University of California Press, pp. 281-297 (Vol. I)

\item[] K. Rose, E. Gurewitz and G. C. Fox, Statistical Mechanics and Phase Transitions in Clustering. Phys. Rev. Lett. 65 (1990) 945.

\item[] V. Roth, T. Lange, M. Braun and J. M. Buhmann,  A Resampling Approach to Cluster Validation. In {\it Proceedings in Computational Statistics: 15th Symposium, Berlin , Germany 2002 (COMPSTAT2002)}, Eds.: W. H\"ardle, Bernd R\"onz (2002) Physica-Verlag, Heidelberg, pp.123-128.

\item[] C. E. Shannon, A mathematical theory of communication. Bell System Tech. J. 27, (1948). 379--423, 623--656. See also: C. Shannon and W. Weaver, {\it The Mathematical Theory of Communication} (1963) University of Illinois Press

\item[] P. Smyth, Model selection for probabilistic clustering using cross-validated likelihood. Statistics and Computing 10 (2000) 63.

\item[] S. Still, W. Bialek and L. Bottou, Geometric Clustering using the Information Bottleneck method. {\it Advances In Neural Information Processing Systems 16} Eds.: S. Thrun, L. Saul, and B. Sch\"olkopf (2004) MIT Press, Cambridge, MA.

\item[] M. Stone, Cross-validatory choice and assessment of statistical predictions. J. R. Stat. Soc. 36 (1974) 111.

\item[] R. Tibshirani, G. Walther and T. Hastie, Estimating the number of clusters in a dataset via the Gap statistic. J. R. Stat. Soc. {\bf B} 63 (2001) 411.

\item[] N. Tishby, F. Pereira and W. Bialek, The information bottleneck method. In {\it Proc. 37th Annual Allerton Conf.} Eds.: B. Hajek and R. S. Sreenivas (1999) University of Illinois, physics/0004057

\item[] A. Treves and S. Panzeri, The upward bias in measures of information derived from limited data samples. Neural Comp. 7 (1995) 399.

\item[] J. H. Ward, Hierarchical groupings to optimize an objective function. J. Am. Stat. Assoc. 58 (1963) 236.

\end{itemize}
\end{document}